\documentclass{aip-cp}
\usepackage[numbers]{natbib}
\usepackage{rotating}
\usepackage{graphicx}

\newtheorem{theorem}{Theorem}

\begin{document}
\def\nn{\nonumber}
\def\deg{\mathop{\rm deg}\nolimits}
\def\str{\mathop{\rm str}\nolimits}
\def\ch{\mathop{\rm char}\nolimits}
\def\modulo{\mathop{\rm mod}\nolimits}
\def\qdots{\mathinner{\mkern1mu\raise1pt\vbox{\kern7pt\hbox{.}}\mkern2mu
 \raise4pt\hbox{.}\mkern2mu\raise7pt\hbox{.}\mkern1mu}}

\def\lb{[\![}
\def\rb{]\!]}
\title{Generalized Quantum Statistics and Lie (Super)Algebras}

\author[]{N.I. Stoilova}
\eaddress[url]{http://www.aip.org}
\eaddress{stoilova@inrne.bas.bg}

\affil{Institute for Nuclear Research and Nuclear Energy, Bould. Tsarigradsko Chaussee 72, 1784 Sofia, Bulgaria }
\maketitle

\begin{abstract}
Generalized quantum statistics, such as paraboson and parafermion statistics, are characterized by triple relations which are related to Lie (super)algebras of type B. The correspondence of the Fock spaces of parabosons, parafermions as well as the Fock space of a system of parafermions and parabosons  to  irreducible representations of (super)algebras of type $B$ will be pointed out. Example of generalized quantum statistics connected to the basic classical Lie superalgebra $B(1|1)\equiv osp(3|2)$ with interesting physical properties, such as noncommutative coordinates, will be given. Therefore the article focuses on the question, addressed already in 1950 by Wigner: do the equation of motion determine the quantum mechanical commutation relation? 
\end{abstract}

\section{INTRODUCTION}
Two classes of particles, bosons and fermions, are usually considered in quantum mechanics. Our interest is in their generalization and more precisely in certain generalized quantum systems. The latter have interesting mathematical structures and representation theory of Lie (super)algebras plays an important role. The aim of the present paper is further exploration the connection between representation theory and quantum systems.

In 1950, Wigner~\cite{Wigner} asked himself the following question 'Do the equations of motion determine the quantum mechanical commutation relations?'
He has shown on the example of a one-dimensional harmonic oscillator with a Hamiltonian $H=\frac{1}{2}(p^2+r^2) \;\;(m=\omega =\hbar=1)$ that one can 
derive the canonical commutation relations (CCRs) $[p,r]=-i$, assuming that Hamilton's equations $\dot{r}=p,\; \dot{p}=-r$  are identical with the Heisenberg equations $\dot{r}=i[H,r]$ and $\dot{p}=i[H,p]$. 
It is known~\cite{Ehrenfest} that the inverse is true. From the Heisenberg equations and the CCRs one obtains  Hamilton's equations and from Hamilton's equations and the CCRs one derives the Heisenberg equations. As a result the question of Wigner was whether one can generalize the concept of a quantum system. He observed that the Hamilton's equations can be identical to the Heisenberg equations for position and momentum operators, which do not necessary satisfy the CCRs. Nowadays such systems are known as Wigner quantum systems (WQS)~\cite{Palev1}. In most of the cases the geometrical properties of such systems are investigated using representation theory of Lie superalgebras. The work of Wigner~\cite{Wigner} led to the theory of parabosons and parafermions introduced by H.S. Green~\cite{Green} in 1953. Paraboson and parafermion statistics are generalizations of ordinary Bose-Einstein and Fermi-Dirac statistics. Algebraically they can be formulated in terms of generators and relations. The parafermion operators $f_j^\pm$, $j=1,2,\ldots, m$, satisfy
the following triple relations
$[[f_{ j}^{\xi}, f_{ k}^{\eta}], f_{l}^{\epsilon}]=\frac 1 2
(\epsilon -\eta)^2
\delta_{kl} f_{j}^{\xi} -\frac 1 2  (\epsilon -\xi)^2
\delta_{jl}f_{k}^{\eta},  
\label{f-rels}$
where $j,k,l\in \{1,2,\ldots,m\}$ and $\eta, \epsilon, \xi \in\{+,-\}$ (to be interpreted as $+1$ and $-1$
in the algebraic expressions $\epsilon -\xi$ and $\epsilon -\eta$), and can be considered as generating elements of
the orthogonal  Lie algebra $B_n\equiv so(2m+1)$~\cite{Kamefuchi,Ryan}. The 
 paraboson operators $b_j^\pm$, $j=1,2,\ldots, n$,  satisfying
$[\{ b_{ j}^{\xi}, b_{ k}^{\eta}\} , b_{l}^{\epsilon}]= (\epsilon -\xi) \delta_{jl} b_{k}^{\eta} 
 +  (\epsilon -\eta) \delta_{kl}b_{j}^{\xi}, 
\label{b-rels} $
are generating elements of the orthosymplectic Lie superalgebra 
$B(0|n)\equiv  osp(1|2n)$~\cite{Ganchev}. The case of $m$ parafermions $f_j^\pm\equiv c_j^\pm$~(\ref{f-rels}) and $n$ parabosons $b_j^\pm \equiv c_{m+j}^\pm$~(\ref{b-rels}) with the so called relative parafermion relations~\cite{Greenberg} 
\begin{eqnarray}
&& \lb\lb c_{ j}^{\xi}, c_{ k}^{\eta}\rb , c_{l}^{\epsilon}\rb =-2
\delta_{jl}\delta_{\epsilon, -\xi}\epsilon^{\langle l \rangle} 
(-1)^{\langle k \rangle \langle l \rangle }
c_{k}^{\eta} +2  \epsilon^{\langle l \rangle }
\delta_{kl}\delta_{\epsilon, -\eta}
c_{j}^{\xi},  \label{para}
\end{eqnarray} 
where
\begin{equation}
\lb a, b \rb = ab-(-1)^{{\deg(a)
\deg(b)}}ba \;\;  \rm{and } \;\; \deg(c_i^\pm)\equiv \langle i\rangle= \left\{ \begin{array}{lll}
 {0} & \hbox{if} & j=1,\ldots ,m \\ 
 {1} & \hbox{if} & j=m+1,\ldots ,n+m
 \end{array}\right.
\end{equation}
 lead to the result that they generate the orthosymplectic Lie superalgebra $B(m|n)\equiv osp(2m+1|2n)$~\cite{Palev2}. As a consequence the parastatistics Fock spaces denoted by $V(p)$ are unitary infinite-dimensional 
$osp(2m+1|2n)$ representations with lowest weight $[-\frac{p}{2},\ldots, -\frac{p}{2}|\frac{p}{2},\ldots, \frac{p}{2}]$, $p$ being a positive 
integer. An explicit and elegant construction of the latter were given in~\cite{SJ}. The case with $n=0$ corresponds to the parafermion Fock 
spaces~\cite{parafermion} and with $m=0$ to the paraboson Fock spaces~\cite{paraboson}. The conclusion for us is that the physical properties of such generalized quantum systems can be investigated applying the representation theory of Lie (super)algebras. In the present article we will apply the results for one pair of parafermion operators $f^\pm\equiv c_1^\pm$ and one pair of paraboson operators $b^\pm \equiv c_{2}^\pm$ generating the Lie superalgebra $osp(3|2)$. Using their parastatistics Fock spaces we will consider and investigate  a 3D harmonic oscillator as a WQS.

\section{The Lie superalgebras $osp(3|2)$ and corresponding to it parastatistics Fock spaces $V(p)$}
The Lie superalgebra 
$osp(3|2)$~\cite{Kac}  consists of matrices of the form
\begin{equation}
\left(\begin{array}{ccccc} a&0&b&x&u  \\
0&-a&c&y&v\\
-c&-b&0&z&w\\
v&u&w&d&e\\
-y&-x&-z&f&-d
\end{array}\right),
\label{osp}
\end{equation}
where the nonzero entries are arbitrary complex numbers.
The even subalgebra $so(3) \oplus sp(2)$ consists of all
matrices~(\ref{osp}), for which $x=y=z=u=v=w=0$, whereas the odd
subspace is obtained taking $a=b=c=d=e=f=0$. 
Let $e_{ij}$ be a 5-by-5 matrix
with 1 on the cross of the $i^{th}$ row and the $j^{th}$
column and zero elsewhere. 
The Cartan subalgebra $H$ of $osp(3|2)$ is the
subspace of diagonal matrices  with basis $h_1=e_{11}-e_{22}, \;$ 
$h_{2}=e_{44}-e_{55}.$ 
It is easy to verify that the matrices 
\begin{eqnarray}
c_{1}^+=f^+= \sqrt{2}(e_{13}-e_{32}), \quad
c_{1}^-=f^-= \sqrt{2}(e_{31}-e_{23}), \\
c_{2}^+=b^+= \sqrt{2}(e_{35}+e_{43}), \quad
c_{2}^-=b^-= \sqrt{2}(e_{34}-e_{53}), 
\label{pb-as-e}
\end{eqnarray}
satisfy the  triple relations~(\ref{para}).
Moreover, the following holds
\begin{theorem}~\cite{Palev2}
As a Lie superalgebra defined by generators and relations, 
$osp(3|2)$ is generated by  elements $c_j^\pm, \; j=1,2$  subject to the parastatistics relations~(\ref{para}).
\end{theorem}
It is straightforward to see that
$
[c_1^-,c_1^+]=-2 h_1, \;  {\rm and} \;\; \{c_{2}^-,c_{2}^+]=2 h_{2}.
$
The parastatistics  Fock space $V(p)$ is defined by:
\begin{eqnarray}
\langle 0|0\rangle=1, \qquad c_j^- |0\rangle = 0, \qquad (c_j^\pm)^\dagger = c_j^\mp,\qquad
\lb c_j^-,c_k^+ \rb |0\rangle = p\delta_{jk}\, |0\rangle , \;\; p=1,2,\ldots \nn
\end{eqnarray}
and 
therefore it is the unitary irreducible representation of
$osp(3|2)$ with lowest weight $(-\frac{p}{2},|\frac{p}{2})$.
\begin{theorem}~\cite{SJ}
An orthonormal basis for the parastatistics Fock space $V(p)$ of one pair of parafermions and one pair of 
parabosons is given by the vectors 
\begin{equation}
|\mu) = 
 \left| \begin{array}{l} \mu_{12}, \mu_{22} \\ \mu_{11} \end{array} \right),\;
{\rm for} \;\mu_{22}=0 \; \mu_{12}=0,1\ldots,p;  {\rm for}\; \mu_{22}=1,2,\ldots \; \mu_{12}=1,2,\ldots,p;  
\mu_{12}-\mu_{11}=\theta=0,1 ({\rm if}\;  \mu_{12}=0, \theta=0). \label{cond}
\end{equation}
The action of the Cartan algebra elements of $osp(3|2)$ is:
\begin{equation}
h_{1}|\mu)=\left(-\frac{p}{2}+\mu_{11}\right)|\mu), 
\quad h_{2}|\mu)=\left(\frac{p}{2}+\mu_{12}+\mu_{22}-\mu_{11}\right)|\mu). \label{h_k} \\
\end{equation}
For the action of the parastatistics operators $c_j^\pm, \; j=1,2$ we have:
\begin{eqnarray}
c^+_1 \left| \begin{array}{l} \mu_{12}, \mu_{22} \\ \mu_{11} \end{array} \right)  
&=& \left(\frac{\mu_{12}+\mu_{22}}{\mu_{12}+\mu_{22}+1}\right)^{\frac{\theta}{2}}{G}_1(\mu_{12},\mu_{22}) 
\left| \begin{array}{l} \mu_{12}+1, \mu_{22} \\ \mu_{11}+1 \end{array} \right)
 \nn\\
&-&\theta\left(\frac{1}{\mu_{12}+\mu_{22}+1}\right) {G}_2(\mu_{12},\mu_{22}) 
\left| \begin{array}{l} \mu_{12},\mu_{22}+1 \\ \mu_{11}+1 \end{array} \right),\label{c1+}\\
c^+_2 \left| \begin{array}{l} \mu_{12}, \mu_{22} \\ \mu_{11} \end{array} \right)  
&=& (1-\theta)\left(\frac{1}{\mu_{12}+\mu_{22}+1}\right)^{\frac{1}{2}}{G}_1(\mu_{12},\mu_{22}) 
\left| \begin{array}{l} \mu_{12}+1, \mu_{22} \\ \mu_{11} \end{array} \right)
 \nn\\
&+&(-1)^{\theta}\left(\frac{\mu_{12}+\mu_{22}}{\mu_{11}+\mu_{22}+1}\right)^{\frac{1}{2}} {G}_2(\mu_{12},\mu_{22}) 
\left| \begin{array}{l} \mu_{12},\mu_{22}+1 \\ \mu_{11} \end{array} \right),\label{c2+}
\end{eqnarray}
\begin{eqnarray}
c^-_1 \left| \begin{array}{l} \mu_{12}, \mu_{22} \\ \mu_{11} \end{array} \right)  
&=& \left(\frac{\mu_{12}+\mu_{22}-1}{\mu_{12}+\mu_{22}}\right)^{\frac{\theta}{2}}{G}_1(\mu_{12}-1,\mu_{22}) 
\left| \begin{array}{l} \mu_{12}-1, \mu_{22} \\ \mu_{11}-1 \end{array} \right)
 \nn\\
&-&(1-\theta)\left(\frac{1}{\mu_{12}+\mu_{22}}\right) {G}_2(\mu_{12},\mu_{22}-1) 
\left| \begin{array}{l} \mu_{12},\mu_{22}-1 \\ \mu_{11}-1 \end{array} \right),\label{c1-}\\
c^-_2 \left| \begin{array}{l} \mu_{12}, \mu_{22} \\ \mu_{11} \end{array} \right)  
&=& \theta\left(\frac{1}{\mu_{12}+\mu_{22}}\right)^{\frac{1}{2}}{G}_1(\mu_{12}-1,\mu_{22}) 
\left| \begin{array}{l} \mu_{12}-1, \mu_{22} \\ \mu_{11} \end{array} \right)
 \nn\\
&+&(-1)^{\theta}\left(\frac{\mu_{12}+\mu_{22}-1}{\mu_{11}+\mu_{22}}\right)^{\frac{1}{2}} {G}_2(\mu_{12},\mu_{22}-1) 
\left| \begin{array}{l} \mu_{12},\mu_{22}-1 \\ \mu_{11} \end{array} \right),\label{c2-}
\end{eqnarray}
\begin{equation}
G_1(\mu_{12},\mu_{22})=\sqrt{ 
\frac{\mu_{12}(\mu_{12}+\mu_{22}+1)(p-\mu_{12})}{\mu_{12}+\mu_{22}+1-{\cal O}_{\mu_{22}+1}}}, 
G_2(\mu_{12},\mu_{22})=\sqrt{ 
\frac{({\cal O}_{\mu_{22}}\mu_{22}+1)({\cal E}_{\mu_{22}+1}(p+\mu_{22})+1)({\cal O}_{\mu_{22}+1}(\mu_{12}+\mu_{22})+1)}
{{\cal E}_{\mu_{22}+1}(\mu_{12}+\mu_{22}-1)+1}}, 
\end{equation}
\begin{equation}
 {\cal E}_{j}=1 \hbox{ if } j \hbox{ is even and 0 otherwise},\quad
 {\cal O}_{j}=1 \hbox{ if } j \hbox{ is odd and 0 otherwise}. \label{EO}
\end{equation}
\end{theorem}
 
\section{Wigner quantum oscillator and $osp(3|2)$}

Consider a three-dimensional harmonic oscillator, namely a quantum system with a Hamiltonian
$
H={{\bf p}^2\over 2m}+ {m \omega^2 \over 2}{\bf r}^2
$
as a Wigner quantum system. This means that we must find the unknown operators  
{\bf r}=$(r_1, r_2, r_3)$ and 
{\bf p}=$(p_1, p_2, p_3)$ so that the following conditions

(i) The state space of the oscillator $W$ is a Hilbert
    space. The physical observables are Hermitian 
    operators in $W$.

(ii) Hamilton's equations $\dot{\bf p}=-m \omega^2 {\bf r},  \dot{\bf r}={{\bf
p}\over m} $ and the Heisenberg
     equations $\dot {\bf p}=-{i\over \hbar}[{\bf p},H],  
\dot {\bf r}=-{i\over \hbar}[{\bf r},H],  $ are identical (as operator equations) in
     $W$.

(iii) The projections of the angular momentum 
      {\bf M}=$(M_1, M_2, M_3)$ are in the
      enveloping algebra of the position 
      {\bf r}=$(r_1, r_2, r_3)$ and  momentum 
      {\bf p}=$(p_1, p_2, p_3)$ operators. Each $M_k$ is linear
      in $(r_1, r_2, r_3)$ and  $(p_1, p_2, p_3)$,
      so that {\bf M}, {\bf r} and {\bf p} transform as
      vectors:
      $[M_j,c_k]=i \sum_{l=1}^3 \varepsilon_{jkl}c_l, 
             c_k=M_k,r_k,p_k,\; j,k=1,2,3 $

hold.

Introduce new
unknown operators
$a_k^\pm=\sqrt{m \omega \over 2 \hbar} r_k \mp
{i \over \sqrt {2m \omega \hbar}}p_k, k=1,2,3. $  
In terms of $a_k^\pm$ the Hamiltonian 
reads
$H={1 \over 2}\omega \hbar \sum_{k=1}^3 \{a_k^+, a_k^- \}$
and the condition (ii) yields ($k=1,2,3$):
\begin{equation}
\sum_{i=1}^3 [ \{a_i^+,a_i^- \},a_k^\pm]=\pm 2a_k^\pm. \label{CCs}
\end{equation}
Let 
\begin{equation}
a_1^\pm={1\over {2 \sqrt{3}}}[c_1^- -c_1^+, c_2^\pm]
\quad a_2^\pm={i\over {2 \sqrt{3}}}[c_1^- +c_1^+, c_2^\pm],
\quad a_3^\pm={1\over {\sqrt{3}}}c_2^\pm. \label{CAO} 
\end{equation}
It is straightforward to check that the operators~(\ref{CAO}) 
satisfy the relations~(\ref{CCs}). Therefore condition (ii) holds.
The projections $(M_1, M_2, M_3)$ of the angular momentum can be defined as
$M_i=-{3\over 4\hbar}\sum_{j,k=1}^3 \varepsilon_{ijk}
\{r_j,p_k\},  i=1,2,3 $ or equivalently 
$M_1={1\over 2}(c_1^+ +c_1^-), 
M_2=-{i\over 2}(c_1^+ -c_1^-), 
M_3={1\over 2}[c_1^-,c_1^+] $ and it is easy to check that condition (iii) is fulfilled. 

Let $W$ be the $osp(3|2)$ module $V(p)$. In each such module $(c_i^-)^\dagger = c_i^+, i=1,2$. 
Therefore the position and the momentum operators are Hermitian
operators (hence also the Hamiltonian $H$, the square of
the angular momentum  ${\bf M}^2$ and its projections
$M_1$, $M_2$, $M_3$ are Hermitian operators) and also condition (i) holds. 
From~(\ref{CAO}) and Theorem 2 one immediately derives that
\begin{equation}
H|\mu) = \frac{\hbar\omega}{2}\{c_2^+,c_2^-\}|\mu)=\frac{\hbar\omega}{2}(p+2\mu_{22}+2\theta)|\mu)
\end{equation}
Thus, the energy spectrum $E_n, n=0,1,2,\ldots$ of the Wigner quantum oscillator is given by
\begin{equation}
E_n={\hbar}\omega(n+\frac{p}{2}), \; n=0,1,2,\ldots .
\end{equation}
The case $p=1$ corresponds to anticommuting pairs of Bose and Fermi operators and the energy spectrum is the same as 
for the one-dimensional canonical harmonic oscillator.

The different coordinates (resp. the different momenta) of the oscillator anticommute
\begin{equation}
\{ r_i, r_j \}=\{ p_i, p_j \}=0, \;\; i\neq j=1,2,3.
\end{equation}
Consequently the Wigner quantum oscillator has neither coordinate nor momentum representation. 
The coordinates, the momenta and the angular momenta operators are on the same footing: the different components do not
commute with each other:
\begin{equation}
[ r_i, r_j ]\neq 0, \; [ p_i, p_j] \neq 0, \;[ M_i, M_j] \neq 0, \; i\neq j=1,2,3.
\end{equation}
The geometry of the oscillator is noncommutative.

Each state $|\mu)$ is an eigenvector of $M_3$:
\begin{equation}
M_3|\mu)=(\frac{p}{2}-2\mu_{12})|\mu), \quad \mu_{12}\leq p.
\end{equation} 
Consequently the Wigner oscillator has an angular momentum $M=\frac{p}{2}, p=1,2,\ldots.$

\section{ACKNOWLEDGMENTS}
The author was supported by  Bulgarian NSF grant DFNI T02/6.
 

\end{document}